\documentclass{article}

\PassOptionsToPackage{numbers,compress}{natbib}

\usepackage[dblblindworkshop, final]{neurips_2026}

\usepackage[utf8]{inputenc}
\usepackage[T1]{fontenc}
\usepackage{hyperref}
\usepackage{url}
\usepackage{booktabs}
\usepackage{amsfonts}
\usepackage{amsmath,amssymb}
\usepackage{nicefrac}
\usepackage{microtype}
\usepackage{xcolor}
\usepackage{graphicx}
\usepackage{placeins}
\usepackage{float}

\newcommand{\repourl}{https://github.com/Shrivatsa-Deshmukh/sbi-gate-vs-ablation}

\title{Gate-Based Feature Importance in Simulation-Based
Inference for EEG: A Cautionary Tale}
\workshoptitle{Sim2Science and Interp4Discovery}

\author{%
  Shrivatsa Deshmukh$^{1,\dagger}$ \quad
  Sasha Brenner$^{2,3}$ \quad
  Thomas R. Kn\"{o}sche$^{2,*}$ \quad
  Nico Scherf$^{2,4,*}$ \\
  \normalfont\footnotesize
  \texttt{deshm080@umn.edu} \quad
  \texttt{brenner@cbs.mpg.de} \quad
  \texttt{knoesche@cbs.mpg.de} \quad
  \texttt{nscherf@cbs.mpg.de} \\[2pt]
  \normalfont\small
  $^{1}$University of Minnesota -- Twin Cities, USA \\
  \normalfont\small
  $^{2}$Max Planck Institute for Human Cognitive and Brain Sciences, Leipzig, Germany \\
  \normalfont\small
  $^{3}$Leipzig University, Germany \quad
  $^{4}$ScaDS.AI, Dresden/Leipzig, Germany
}

\begin{document}

\maketitle

{\renewcommand{\thefootnote}{}\footnotetext{$^{\dagger}$Lead and corresponding author. \quad $^{*}$Equal senior contribution.}}

\begin{abstract}
Simulation-based inference (SBI) provides a powerful way to fit mechanistic models to data, but offers little insight into which information in the data shapes the inferred parameter values. It could therefore benefit from robust and inexpensive interpretability methods. Learned weighting mechanisms such as feature gates provide a potential route to low-cost interpretability, although their interpretation assumes that predictive training reveals which inputs drive performance. We examined this assumption in a computational neuroscience setting, where neural posterior estimation (NPE) is used to infer four physiological parameters of a dynamical model of interacting neuronal populations (Jansen-Rit). We tested whether the learned gate weights captured feature importance by comparing them with a retrain-based ablation benchmark, in which each feature was removed in turn and the NPE retrained. We considered two jointly trained gates: independent sigmoid weights and competing softmax weights. If these reliably reflect feature importance, they should align with the ablation-based estimates. Across nine independent training seeds, however, gate-based and ablation-based importance were negatively correlated (Spearman's rho approximately -0.61 for sigmoid gating and -0.56 for softmax gating). The feature that ablation identified as most important, total signal power, was among the features the gate down-weighted most, despite gating costing negligible predictive accuracy. These results extend a pattern previously documented for attention mechanisms in natural language processing and for explainability methods in electroencephalography (EEG) deep learning to the simulation-based inference setting, and suggest caution when interpreting learned gate weights as measures of causal feature importance in scientific machine-learning pipelines.
\end{abstract}

\section{Introduction}

Mechanistic models of neural activity, from single-compartment Hodgkin-Huxley
neurons to neural mass models of cortical columns, offer parameters that are
biologically interpretable in a way purely statistical decoders do not provide. The
Jansen--Rit model \citep{jansen1995electroencephalogram} is a canonical case: a
six-state neural mass model of a cortical column that reproduces electroencephalography (EEG) rhythms from a
small number of physiologically grounded parameters. Fitting such models to data
remains hard \citep{hartoyo2019parameter,caccamo2026dynamics,hadida2018bayesian}. Their dynamics are nonlinear, and the likelihood of the observed data
given the model's parameters $p(x \mid \theta)$ has no closed form, since it would
require marginalizing over the full space of stochastic input trajectories
consistent with the observed output \citep{sisson2018handbook}. Classical approaches, including grid search,
moment matching, and approximate Bayesian computation, scale poorly as the number
of free parameters grows \citep{cranmer2020frontier}.

Simulation-based inference (SBI) completely sidesteps evaluating intractable likelihoods by training a neural density estimator
directly on simulation outputs, learning an amortized mapping from data features to a
posterior over parameters
\citep{cranmer2020frontier,papamakarios2016fast,greenberg2019automatic}.
Gon{\c{c}}alves et al.~\citep{goncalves2020training} established this approach for
mechanistic neuroscience models, applying neural posterior estimation to single-neuron and small-circuit models, and showed that hand-designed
summary-statistic embeddings are often sufficient for accurate, well-calibrated
inference.

In scientific applications of SBI, it becomes crucial to understand how each feature of the data contributes to parameter inference. Normally, a feature's contribution is established only after training;
the gold standard in SBI is retrain-based ablation, in which the model is
retrained with one feature excluded and the change in the inferred posterior is quantified, a causal measure by construction. This is the procedure the current
practice guide points to for assessing which features constrain parameters
\citep{deistler2025simulation}, and the criterion against which Beck et
al.~\citep{beck2022efficient} validated a faster analytic alternative, Feature
Selection through Likelihood Marginalization.
Both measure the effect that excluding a feature actually has on the posterior:
one by performing the exclusion directly, the other by approximating it analytically.
An appealing, cheaper alternative is to learn the weighting itself by
incorporating a small differentiable gate, trained end-to-end with the rest of the
network. The gate multiplies each feature by a weight in $(0,1)$, and after
training it can be read as importance at no extra cost.
Indeed, gates of exactly this form are reported in various settings: as electrode importance in EEG deep
learning \citep{tao2023eeg,tong2023learnable}; and, in tabular learning and
forecasting, the reading is built into the methods themselves, through learned input
masks \citep{arik2021tabnet}, attention-based feature selection \citep{gui2019afs},
and variable-selection layers \citep{lim2021temporal}. In SBI, however, the idea appears not
to have been taken up (although it has been suggested \citep{kakadia2025vision}), and we believe it could be attractive in principle, since
the standard ablation test costs one full retraining per feature.

\begin{figure}[t]
\centering
\includegraphics[width=\linewidth]{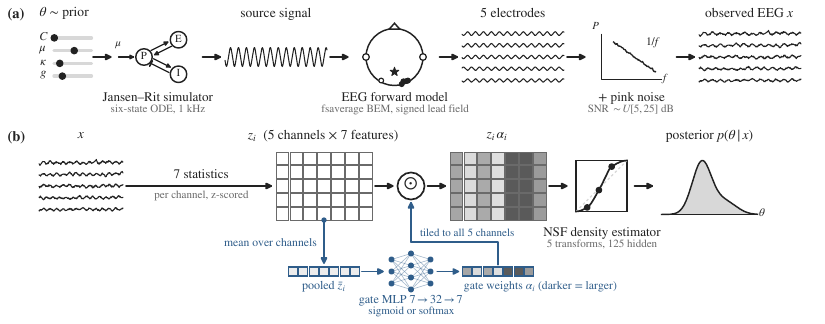}
\caption{\textbf{Pipeline overview.} (a)~Simulation: parameters $\theta$ drawn from
the prior drive the Jansen--Rit model, giving the observed EEG $x$. (b)~Inference:
seven z-scored summary statistics per channel form the $5\times7$ array $z_i$; in
gated configurations (blue) a small MLP produces one weight $\alpha_i$ per feature,
tiled to all five channels, so the estimator receives $z_i\,\alpha_i$ in place of
$z_i$. Shades in (b): seed-averaged sigmoid importance of
Table~\ref{tab:importance}.}
\label{fig:pipeline}
\end{figure}

Evidence regarding the interpretability of learnable gates in deep-learning pipelines is still mixed, however.
On the one hand, architectures built for feature selection do sometimes recover known relevant variables on synthetic data.
For example, TabNet's sparse masks identified ground-truth features \citep{arik2021tabnet}; importantly, however, their weights are not generally guaranteed to represent importance.
More negative evidence comes from the UCI Adult dataset, where learned mask and
attention scores showed little agreement with KernelSHAP, a method that estimates
each feature's contribution to a prediction \citep{borisov2024deep}. Results
from natural language processing (NLP) are similarly cautionary: attention weights sometimes tracked input
effects, but too inconsistently to provide reliable importance measures
\citep{serrano2019attention}.
In EEG, explanation quality also depends on the method: saliency failed a model-randomization check, whereas DeepLIFT did not \citep{ravindran2023empirical,pmlr-v70-shrikumar17a}.
More fundamentally, jointly trained selectors can encode predictions in their masks \citep{jethani2021have}, while unconstrained multiplicative scales may be non-identifiable under downstream rescaling \citep{liu2017learning}.
Taken together, these findings suggest that learned weights can be informative
without necessarily providing reliable importance estimates. That uncertainty extends to SBI, where the relationship between learned weights and retrain-based ablation remains largely unexplored.

We compare the gating and ablation measures of importance directly in a Jansen--Rit EEG model (Figure~\ref{fig:pipeline}).
Neural posterior estimators (NPEs) are trained over four mechanistic parameters: connectivity
strength, input drive, the joint excitatory/inhibitory timescale-scaling factor,
and inhibitory synaptic gain. These parameters are inferred from simulated EEG,
using seven statistical-summary features computed at five electrodes. Two gated
variants, independent (sigmoid) and competing (softmax) per-feature weighting, are
compared against an ungated baseline in terms of model performance, and against the gold-standard measure of
feature importance, obtained by retraining the full pipeline from scratch with each
feature excluded in turn and measuring the resulting drop in posterior accuracy. We only use simulated EEG data in order to have a clean ground truth to evaluate the feasibility of the gating method. Indeed, many more challenges are introduced by real data which are beyond the scope of this work.\footnote{All code required to reproduce these results is available at
\url{\repourl}; its contents are
described in Appendix~\ref{app:code}.}

\section{Methods}

\subsection{Simulation and forward model}
\label{sec:sim}

We simulated EEG using the standard six-state Jansen--Rit neural mass model of a
cortical column \citep{jansen1995electroencephalogram} (Figure~\ref{fig:pipeline}a),
integrated with a fixed-step Euler scheme at 1000~Hz. We varied and subsequently
inferred four parameters: connectivity strength ($C$), mean input drive ($\mu$), a
joint scaling factor for the excitatory and inhibitory postsynaptic time constants
($\kappa$), and inhibitory synaptic gain ($g$). $\kappa$ scales both time constants
proportionally and holds their ratio fixed at 2:1; these time constants are what set
the model's oscillation frequency \citep{david2003neural}. Excitatory gain
was held fixed deliberately, so that the excitatory-inhibitory gain degeneracy
reported for larger Jansen--Rit parameter sets \citep{tilwani2025deep} cannot arise
here by construction. All four parameters were given independent log-uniform priors,
enforcing positivity and giving roughly equal prior weight across their plausible
range; the ranges are listed in Appendix~\ref{app:sim}.

We projected the single-source output to scalp-sensor space using
MNE-Python's \citep{gramfort2013meg,gramfort2014mne} fsaverage template head model
and a three-layer boundary-element (BEM) forward solution, with the source located
in the visual cortex. From the resulting lead-field, we selected the five electrodes
with the highest absolute sensitivity to this source, then projected the source
signal using each electrode's actual (signed) lead-field value rather than its
magnitude (Figure~\ref{fig:pipeline}a), preserving realistic scalp topography rather
than collapsing to a single scalar channel. We added pink (1/f) noise to each
simulated trial, with signal-to-noise ratio (SNR) drawn independently per trial from a
uniform distribution over 5--25~dB during training, a range specifically chosen to
span moderate to light signal corruption. Simulation and preprocessing parameters
are given in full in Appendix~\ref{app:sim}.

\subsection{Feature embedding and gating}
\label{sec:gating}

Figure~\ref{fig:pipeline}b illustrates the embedding and gating. From each of the
five electrode channels, we computed seven hand-crafted summary statistics:
skewness, kurtosis, spectral slope, total signal power, dominant frequency, Hjorth
mobility, and Hjorth complexity \citep{hjorth1970eeg}. In gated configurations, each
feature value was first z-scored per feature and channel using the training-set
statistics, giving $z_i$ for feature $i$.\footnote{The notation $z_i$ omits channel dependence.} A small multilayer perceptron mapped the
seven features, pooled (averaged) across channels, to seven per-feature, sample-dependent gate weights
$\alpha_i \in (0,1)$. These were then tiled back out across channels and applied
multiplicatively, so that the density estimator received $z_i\,\alpha_i$ in place of $z_i$.
Note that all five channels of feature $i$ shared the same weight $\alpha_i$. Appendix~\ref{app:gating} discusses the design choice.

We compared two gate activation functions: an element-wise sigmoid and a softmax.
The sigmoid weighted each feature independently with no explicit temperature term,
whereas softmax introduced cross-feature competition by forcing the seven weights to sum to one.
The gate parameters were optimized jointly with the rest of the
network on the same objective. Crucially, no separate loss or gradient-blocking was applied to
its forward pass.

The per-feature importance we report for the gate (Table~\ref{tab:importance}) is
$|z_i| \cdot \alpha_i$, averaged over samples, channels, and seeds. Because the
estimator's input is the signed product, we took the absolute value of $z_i$ before
averaging so that magnitude contributions of either sign would not cancel across the
evaluation set.
We also report a robustness check using the raw gate weight alone in the Results section.

\subsection{Inference, ablation, and evaluation}
\label{sec:inference}

We estimated the posterior over the Jansen--Rit parameters given the feature
embedding using single-round neural posterior estimation as implemented in the
\texttt{sbi} package \citep{tejerocantero2020sbi}, with a neural spline flow
\citep{durkan2019neural} density estimator. Every configuration was trained
independently across nine seeds, with training parameter sets redrawn independently
for each seed; architecture and training hyperparameters are given in
Appendix~\ref{app:training}.

We estimated feature importance by retrain-based leave-one-feature-out ablation: for
each of the seven features and each seed, we retrained the full pipeline from
scratch with that feature excluded from the embedding, using the same simulated
training set as the corresponding full-feature run for that seed. We quantified importance as the drop in mean $R^2$ relative to the full-feature model trained on the same seed, where $R^2$ is the coefficient of determination between true and posterior-mean parameter values, computed on a held-out set for each of the four parameters and then averaged over them. We used this retrain-based procedure, rather than a
post-hoc masking heuristic, as it measures each feature's causal contribution by
construction.

We computed all accuracy metrics on a fixed set of 250 held-out simulations per
seed, drawn as described in Appendix~\ref{app:training}. For each held-out
simulation we drew 1{,}000
posterior samples and took the posterior mean as the point estimate, from which we
computed $R^2$ and the fraction of 50\%, 90\%, and 95\% credible intervals containing the true value,
following the spirit of simulation-based calibration checks
\citep{talts2018validating}, which we will refer to as credible-interval coverage henceforth. Additional metrics and per-parameter breakdowns are
given in Appendix~\ref{app:extra}. For paired comparisons across seeds, such as
ablation- versus gate-based importance rankings, we used an exact sign-flip test
over all $2^{9}=512$ possible $\pm 1$ sign assignments of the nine per-seed
statistics, avoiding both distributional assumptions and Monte Carlo approximation
error.\footnote{With nine seeds the smallest attainable two-sided $p$-value is
$2/512 \approx 0.0039$. Several results below sit at this floor and should be read
as ``all seeds agreed,'' not as evidence for an arbitrarily small true $p$-value.}

\section{Results and Discussion}

Across nine independent training seeds, ablation identified total signal power as
the most important of the seven features by a wide margin: excluding it dropped mean
$R^2$ by 0.210 (SD 0.009), roughly five times the drop from the next-largest
feature. Testing each feature's own ablation effect against the null with
the exact sign-flip test described in Methods, six of the seven features had an effect
significantly different from zero (Table~\ref{tab:importance}),
spectral slope being the only exception ($p=0.727$).

\begin{table}[htbp]
\centering
\caption{\textbf{Feature importance: gold-standard ablation vs.\ learned gate.}
Mean (SD) drop in $R^2$ across 9 seeds when the feature is excluded and the
pipeline retrained from scratch, with the exact sign-flip test's $p$-value.
$\dagger$ marks the test's minimum attainable value, $p=0.0039$, reached through
unanimous agreement across all nine seeds. Gate columns give the per-feature
importance $|z_i|\,\alpha_i$ of Section~\ref{sec:gating}, pooled across channels. Rank 1 = most important by that method.}

\label{tab:importance}
\small
\setlength{\tabcolsep}{4pt}
\begin{tabular}{lccccccc}
\toprule
& \multicolumn{3}{c}{Ablation} & \multicolumn{2}{c}{Sigmoid gate} & \multicolumn{2}{c}{Softmax gate} \\
\cmidrule(lr){2-4} \cmidrule(lr){5-6} \cmidrule(lr){7-8}
Feature & $R^2$ drop (SD) & $p$ & Rank & $|z_i|\,\alpha_i$ & Rank & $|z_i|\,\alpha_i$ & Rank \\
\midrule
Total signal power     & 0.210 (0.009) & $\dagger$    & 1 & 0.343 & 7 & 0.094 & 6 \\
Kurtosis             & 0.042 (0.010) & $\dagger$    & 2 & 0.349 & 6 & 0.087 & 7 \\
Skewness             & 0.041 (0.012) & $\dagger$    & 3 & 0.451 & 4 & 0.123 & 5 \\
Hjorth complexity    & 0.041 (0.010) & $\dagger$    & 4 & 0.471 & 3 & 0.124 & 4 \\
Hjorth mobility      & 0.031 (0.009) & $\dagger$    & 5 & 0.595 & 1 & 0.126 & 2 \\
Dominant frequency   & 0.012 (0.016) & 0.031        & 6 & 0.594 & 2 & 0.132 & 1 \\
Spectral slope       & 0.002 (0.013) & 0.727 (n.s.) & 7 & 0.443 & 5 & 0.124 & 3 \\
\bottomrule
\end{tabular}
\end{table}

In contrast, the gate did not favor total signal power. In fact, it assigned the feature the lowest or
second-lowest weight under both activations (Table~\ref{tab:importance}), while its
two most-favored features, dominant frequency and Hjorth mobility, had ablation
effects an order of magnitude smaller.

This disagreement was systematic. We computed the Spearman
rank correlation between ablation-based and gate-based importance separately for
each seed (seven features per seed) and tested the seed-averaged correlation against
zero with the exact sign-flip test. Independent (sigmoid) gating gave
$\rho = -0.607$ and competing (softmax) gating gave $\rho = -0.556$, both at
$p=0.0039$. Meanwhile, the two gating variants agreed with each other on which features
mattered ($\rho = 0.643$, $p=0.0039$). Thus, the gate was internally consistent and
reproducible across activation functions; it simply tracked something other than
what ablation tracked.

The negative correlations remained significant after excluding any single feature.
The full exclusion analysis is given in
Appendix~\ref{app:extra}, Table~\ref{tab:gate_correlation}. They also persisted after jointly excluding the top-scoring features of ablation and gating, \textit{i.e.}, total signal power and dominant frequency, respectively (sigmoid: $\rho=-0.300$, $p=0.027$; softmax: $\rho=-0.333$, $p=0.035$; Appendix~\ref{app:extra}).
Moreover, the negative sign of the correlation also significantly persisted using the raw gate weights alone (sigmoid: $\rho=-0.298$, $p=0.0039$; softmax: $\rho=-0.214$, $p=0.008$), showing that it was not purely an artifact of including feature magnitude in the score. It did weaken, which suggests feature magnitude does contribute to the effect.

We next asked whether this disagreement could reflect a poorly performing model. A
priori, it seemed possible that the gate's disagreement with ablation simply
reflected a network that was performing poorly overall.
This was not the case. Gating cost little in
raw accuracy: mean $R^2$ was 0.791 without gating, 0.792 with per-feature (sigmoid) gating,
and 0.761 with competing (softmax) gating, differences small relative to their seed-to-seed
spread. Credible-interval coverage was close to the nominal 50\%, 90\%, and 95\% levels under all three
configurations, thus showing no evidence of systematic over- or under-confidence from either
gating variant. Moreover, ordering the four parameters by per-parameter $R^2$ resulted in an identical ranking under every configuration
(Appendix~\ref{app:extra}).

Our main finding was that ablation-based and gate-based feature importance were negatively correlated for both
gating variants.
This disagreement is arguably more pronounced than a weak correlation, such as, for example, that reported by Jain and Wallace \citep{jain2019attention} in a natural language processing setting.
Our results therefore reinforce the broader conclusion that mechanisms purely optimized for predictive
accuracy should not automatically be interpreted as reliable indicators of feature importance. In the terms of Jacovi and Goldberg \citep{jacovi2020faithfully}, the gate was plausible but not faithful: its feature-importance rankings were consistent across training seeds and moderately similar between the sigmoid and softmax variants ($\rho=0.643$), yet it assigned the least importance to the feature whose removal most impaired performance.

Even if a measure of causal importance were extractable from the network's learned
parameters, we have shown that in this setting it is not as simple as reading gate
values.
What the gate tracked instead is not established by this study, nor is whether the disagreement persists where ground
truth is not simulated and ablation must be scored against a proxy such as held-out
predictive likelihood.

\section{Conclusion}
A feature-gating mechanism trained end-to-end alongside a simulation-based posterior
estimator did not recover the feature importance established by gold-standard
retrain-based ablation, and did so while costing negligible accuracy and being
optimized on exactly the objective the modeler cares about. Practitioners using
learned gating or attention as a cheap stand-in for feature-importance analysis in
scientific, simulation-based inference pipelines should treat this as a caution
rather than a shortcut. A gating mechanism optimized purely for predictive accuracy is not
thereby guaranteed to possess explanatory power.

\vskip\baselineskip

\paragraph{Responsible-use statement.} The risk this work identifies is misplaced
trust: a coherent-looking gate can support an incorrect scientific conclusion under
the appearance, rather than the substance, of mechanistic justification. We
therefore recommend that learned weighting be validated against a causal measure,
such as the retrain-based ablation used here, before it is reported as importance.

\begin{ack}
The authors declare no competing interests.
\end{ack}

\bibliography{references}

\clearpage
\appendix
\setcounter{table}{0}
\renewcommand{\thetable}{D\arabic{table}}

\section{Simulation details}
\label{app:sim}

In Section~\ref{sec:sim}, all four inferred parameters were drawn from independent
log-uniform priors across the following ranges: $C \in [135, 270]$, $\mu \in [120, 350]$,
$\kappa \in [0.75, 1.25]$, and $g \in [0.5, 2.0]$. Input noise
standard deviation ($\sigma$) was fixed at 4.5 rather than inferred, since once observation
noise is added downstream, $\sigma$ and observation noise are not jointly
identifiable from the signal alone. Each simulation discarded an initial 2~s
transient and retained the following 3~s, decimated from the internal 1000~Hz
integration rate to a 250~Hz output rate. Training parameter sets were drawn using
scrambled Sobol sequences, for better prior coverage than independent sampling would
give, redrawn independently for each of the nine training seeds used throughout.

The signed lead-field projection described in Section~\ref{sec:sim} preserves sign
inversion across electrode pairs, a feature of realistic scalp topography that
a magnitude-only projection would discard. The per-trial SNR draw over 5--25~dB
ensures that the density estimator learns a posterior robust to a spread of data
quality rather than overfitting to a single noise level.

\section{Gating design details}
\label{app:gating}

In Section~\ref{sec:gating}, one gate weight $\alpha_i$ was applied to all five channels of
feature $i$, rather than learning a separate weight for each of the thirty-five
(feature, channel) pairs. We made this choice so the gate and the ablation
procedure would operate on the same unit of analysis: ablation excludes a feature
across all five channels at once, never a single channel of a single feature, so a
gate that weighted channels independently would not be answering the same question.
Table~\ref{tab:importance}'s two importance columns are comparable by construction
because both are computed over that same seven-feature unit.

\section{Training and evaluation details}
\label{app:training}

In Section~\ref{sec:inference}, we estimated the posterior with a neural spline flow using five transform layers with 125 hidden features per layer. We trained on 32{,}768 simulations by default, with batch size
512, learning rate $5\times10^{-4}$, for up to 250 epochs with early stopping after
15 epochs without validation-loss improvement, and a 15\% validation split.

The held-out evaluation set of 250 simulations was drawn once per seed and reused
across every configuration and ablation condition, so comparisons were not affected
by which simulations happened to be drawn. Normalized RMSE (NRMSE) is
root-mean-squared error divided by each parameter's prior range, which makes error
comparable across parameters despite their different prior scales. Credible-interval
coverage at a given level is the fraction of held-out simulations for which the true
parameter value fell within the corresponding symmetric, equal-tailed percentile
interval of the 1{,}000 posterior samples drawn for that simulation.

All experiments were run on a single consumer NVIDIA RTX 4050 Laptop GPU. The 90 training runs reported here comprised three gating
configurations and seven leave-one-feature-out ablation conditions across nine
seeds each, and took approximately 5 GPU-hours in
total, or roughly 3~minutes per run including simulation, training, and
evaluation.

\section{Exclusion analysis, accuracy, and per-parameter recovery}
\label{app:extra}

Table~\ref{tab:gate_correlation} gives the full single-exclusion rank correlations
summarized in Results; every column is significant at $p=0.0039$.
Table~\ref{tab:accuracy} gives pooled accuracy and calibration by gating
configuration. Tables~\ref{tab:perparam} and~\ref{tab:perparam_configs} break
recovery down by parameter.

\begin{table}[H]
\centering
\caption{\textbf{Rank correlation between ablation-based and gate-based
importance.} Seed-averaged Spearman $\rho$, computed on all seven features and
recomputed on the six remaining features after excluding the named feature.}
\label{tab:gate_correlation}
\small
\setlength{\tabcolsep}{4pt}
\begin{tabular}{lcccccccc}
\toprule
 & All 7 & Total & Dominant & Hjorth & Hjorth & Kurtosis & Skewness & Spectral \\
 & features & signal power & frequency & mobility & complexity & & & slope \\
\midrule
Sigmoid $\rho$ & $-0.607$ & $-0.410$ & $-0.562$ & $-0.632$ & $-0.683$ & $-0.517$ & $-0.581$ & $-0.771$ \\
Softmax $\rho$ & $-0.556$ & $-0.410$ & $-0.530$ & $-0.511$ & $-0.651$ & $-0.568$ & $-0.562$ & $-0.575$ \\
\bottomrule
\end{tabular}
\end{table}

\begin{table}[H]
\centering
\small
\caption{\textbf{Model accuracy and calibration by gating configuration.} Mean (SD) across 9 seeds.}
\label{tab:accuracy}
\setlength{\tabcolsep}{4pt}
\begin{tabular}{lcccccc}
\toprule
Configuration & $R^2$ & NRMSE & Pearson $r$ & 50\% coverage & 90\% coverage & 95\% coverage \\
\midrule
No gating & 0.79 (0.007) & 0.13 (0.002) & 0.89 (0.004) & 0.49 (0.015) & 0.88 (0.007) & 0.94 (0.009) \\
Sigmoid gating       & 0.79 (0.016) & 0.13 (0.005) & 0.89 (0.009) & 0.50 (0.009) & 0.89 (0.011) & 0.95 (0.008) \\
Softmax gating       & 0.76 (0.019) & 0.14 (0.006) & 0.87 (0.011) & 0.53 (0.015) & 0.90 (0.011) & 0.95 (0.008) \\
\bottomrule
\end{tabular}
\end{table}

\begin{table}[H]
\centering
\caption{\textbf{Per-parameter recovery and calibration under the ungated baseline.}
Mean across 9 seeds.}
\label{tab:perparam}
\small
\setlength{\tabcolsep}{4pt}
\begin{tabular}{lcccccc}
\toprule
Parameter & $R^2$ & NRMSE & Pearson $r$ & 50\% cov. & 90\% cov. & 95\% cov. \\
\midrule
Connectivity ($C$)        & 0.812 & 0.122 & 0.902 & 0.497 & 0.877 & 0.941 \\
Input drive ($\mu$)        & 0.702 & 0.157 & 0.838 & 0.492 & 0.880 & 0.939 \\
Timescale scaling ($\kappa$) & 0.738 & 0.144 & 0.860 & 0.472 & 0.880 & 0.941 \\
Synaptic gain ($g$)        & 0.915 & 0.086 & 0.958 & 0.502 & 0.903 & 0.952 \\
\bottomrule
\end{tabular}
\end{table}

\begin{table}[H]
\centering
\caption{\textbf{Per-parameter recovery ($R^2$) across gating configurations.}
Mean across 9 seeds per configuration.}
\label{tab:perparam_configs}
\small
\setlength{\tabcolsep}{4pt}
\begin{tabular}{lccc}
\toprule
Parameter & No gating & Sigmoid gating & Softmax gating \\
\midrule
Connectivity ($C$)           & 0.812 & 0.813 & 0.778 \\
Input drive ($\mu$)           & 0.702 & 0.695 & 0.653 \\
Timescale scaling ($\kappa$)  & 0.738 & 0.745 & 0.725 \\
Synaptic gain ($g$)           & 0.915 & 0.914 & 0.885 \\
\bottomrule
\end{tabular}
\end{table}

\FloatBarrier

\section{Code, environment, and licenses}
\label{app:code}

The repository linked in the Introduction contains the full pipeline used here: the
Jansen--Rit simulator and forward-model construction of Section~\ref{sec:sim}, the
sigmoid and softmax gate implementations of Section~\ref{sec:gating}, the
retrain-based ablation loop of Section~\ref{sec:inference}, and the evaluation scripts that produce every table in this paper. The exact
commands corresponding to each configuration and ablation condition are listed in
the repository's README.

Results were produced under Python 3.11 with a CUDA 12.4 build of PyTorch v2.5.1. The full pinned environment is given in the repository's
\texttt{requirements.txt}.

All software used is open source. \texttt{sbi} v0.25.0 (github.com/sbi-dev/sbi) is
used under the Apache License 2.0.
MNE-Python v1.11.0 (github.com/mne-tools/mne-python) is used under the BSD 3-Clause
License. PyTorch v2.5.1, NumPy v1.26.4, SciPy
v1.12.0 and pandas v2.3.3 are used under the BSD 3-Clause License, and Matplotlib
v3.10.8 under its own BSD-compatible license.

\end{document}